\documentclass[doc,natbib]{apa6}\usepackage[]{graphicx}\usepackage[table]{xcolor}
\makeatletter
\def\maxwidth{ %
  \ifdim\Gin@nat@width>\linewidth
    \linewidth
  \else
    \Gin@nat@width
  \fi
}
\makeatother

\definecolor{fgcolor}{rgb}{0.345, 0.345, 0.345}
\newcommand{\hlstr}[1]{\textcolor[rgb]{0.192,0.494,0.8}{#1}}%
\newcommand{\hlstd}[1]{\textcolor[rgb]{0.345,0.345,0.345}{#1}}%
\newcommand{\hlkwb}[1]{\textcolor[rgb]{0.69,0.353,0.396}{#1}}%

\usepackage{framed}
\makeatletter
\newenvironment{kframe}{%
 \def\at@end@of@kframe{}%
 \ifinner\ifhmode%
  \def\at@end@of@kframe{\end{minipage}}%
  \begin{minipage}{\columnwidth}%
 \fi\fi%
 \def\FrameCommand##1{\hskip\@totalleftmargin \hskip-\fboxsep
 \colorbox{shadecolor}{##1}\hskip-\fboxsep
     \hskip-\linewidth \hskip-\@totalleftmargin \hskip\columnwidth}%
 \MakeFramed {\advance\hsize-\width
   \@totalleftmargin\z@ \linewidth\hsize
   \@setminipage}}%
 {\par\unskip\endMakeFramed%
 \at@end@of@kframe}
\makeatother

\definecolor{shadecolor}{rgb}{.97, .97, .97}
\definecolor{messagecolor}{rgb}{0, 0, 0}
\definecolor{warningcolor}{rgb}{1, 0, 1}
\definecolor{errorcolor}{rgb}{1, 0, 0}
\newenvironment{knitrout}{}{} 

\usepackage{alltt}
\usepackage{epsfig,amsmath,alltt,bm,url,subcaption,lineno}
\usepackage[english]{babel}
\usepackage[table]{xcolor}

\title{Opaque prior distributions in Bayesian latent variable models}
\fourauthors{Edgar C. Merkle}{Oludare Ariyo}{Sonja D. Winter}{Mauricio Garnier-Villarreal}
\fouraffiliations{University of Missouri, USA\\ https://orcid.org/0000-0001-7158-0653}{University of Essex, UK\\ https://orcid.org/0000-0003-3375-1831}{University of Missouri, USA\\ https://orcid.org/0000-0002-2203-002X}{Vrije Universiteit Amsterdam, The Netherlands\\ https://orcid.org/0000-0002-2951-6647}
\abstract{
  We review common situations in Bayesian latent variable models where the prior distribution that a researcher specifies differs from the prior distribution used during estimation. These situations can arise from the positive definite requirement on correlation matrices, from sign indeterminacy of factor loadings, and from order constraints on threshold parameters. The issue is especially problematic for reproducibility and for model checks that involve prior distributions, including prior predictive assessment and Bayes factors. In these cases, one might be assessing the wrong model, casting doubt on the relevance of the results. The most straightforward solution to the issue sometimes involves use of informative prior distributions. We explore other solutions and make recommendations for practice.

  }

\authornote{This work was supported by the Institute of Education Sciences, U.S. Department of Education, Grant R305D210044. We thank Bob Carpenter and two reviewers for comments that improved the paper. All remaining errors are due to the authors.
  Correspondence to Edgar Merkle, 28A McAlester Hall, University of Missouri, Columbia, MO 65203.
  Email: merklee@missouri.edu.}
\shorttitle{Opaque priors}

\let\proglang=\textsf
\let\pkg=\emph

\newcommand{\redc}{\cellcolor{red!20}}

\IfFileExists{upquote.sty}{\usepackage{upquote}}{}
\begin{document}
\maketitle

\section{Introduction}
Bayesian latent variable models, including structural equation models and time series models, are becoming increasingly complex to accommodate the datasets that modern technology permits \citep[e.g.,][]{aspham18,bue19,ctsem,deplaiyang21,endkel18,garnier_betweenitem_2021,haamer20,haarou21,hoijsch18,kapche22,kelend23,levend21,magnus_multidimensional_2021,mancul22,miolev21,pagpac22,rasmar22,schoot13,silaze22,uan22,uan23,ulipoh22,vanbro21,wiljac21,zyphur21}. To ensure that these complex models are working correctly, it is important to fully understand the problematic issues that can arise in simpler Bayesian models with latent variables. We think that this understanding is not yet fully developed. For example, researchers often overlook the fact that multiple WAIC metrics are available for the same model, depending on whether or not one includes latent variables in the likelihood \citep{merfur19}. And recent evidence indicates that small-variance priors can mask misfit in other parts of the model \citep{wiberg_limited_2023}.  Continuing with the theme of problematic issues, the current paper discusses problems with prior distributions that are easy to overlook and that can lead to incorrect results and model summaries. 


The issues that we consider here can generally be called {\em opaque prior distributions}. They involve the fact that, for some Bayesian models, the prior distribution that a researcher specifies is not the prior distribution that the estimation method actually uses. This happens because prior distributions are influenced by various details surrounding MCMC implementation, beyond the researcher's specified prior. These issues are already known to many statisticians, but they have some unique manifestations in Bayesian psychometric models with latent variables. Further, these issues have not received much attention in the context of Bayesian psychometric modeling.

In psychometric models with latent variables, opaque prior distributions can arise from positive definite constraints associated with model covariance matrices, from sign indeterminacies in factor loadings, and from order constraints associated with threshold parameters. One example was considered by \cite{ghodun09}, who describe a parameter expansion algorithm for estimating Bayesian factor analysis models. They developed a Gibbs sampler for an overparameterized model in which the factor loadings were not identified, then translated the unidentified parameters to identified parameters via postprocessing. Of interest here, they showed that the priors for the identified loadings (obtained via postprocessing) differed from the priors for the unidentified loadings. We consider this example in more detail later.

Opaque prior distributions can cause a variety of problems. One way in which these problems arise is through recent studies where researchers seek to recommend the ``best'' prior distributions for various models. For example, specific prior distributions have been recommended for factor loadings and correlation parameters in confirmatory factor analysis \citep{ludtke_comparison_2021, ulitzsch_alleviating_2021}, factor loadings and intercept parameters in item factor analysis with dichotomous indicators \citep{bainter_bayesian_2017}, regression parameters in mediation models with latent variables \citep{miolev21}, covariance matrices in SEM \citep{liu_understanding_2022, van_zundert_prior_2022}, random effect parameters in multilevel SEMs \citep{vanbro21, zitzmann_performance_2021}, and variance parameters in approximate measurement invariance modeling \citep{pokropek_choosing_2020}. However, as we will show, opaque priors imply that such recommendations can be specific to the software used. In other words, alternative software packages may implement the same model in a different way, leading to a different implied prior distribution even though the user always specified the same prior distribution.

We could also experience problems with metrics that explicitly involve evaluation of the prior, including Bayes factors \citep[e.g.,][]{hecboe22,kasraf95} and prior predictive checks \citep[e.g.,][]{van20}. In both cases, if a researcher specifies a model with opaque prior distributions, then it is easy to use the wrong prior distributions to compute the metrics. Relatedly, some methods for verifying the accuracy of one's MCMC algorithm involve generating data from prior distributions. If researchers use the wrong priors, then they may conclude that their MCMC algorithm is problematic even though it runs correctly, or vice versa.

The intent of this paper is to illustrate opaque prior distributions and to provide solutions for avoiding them. This can allow for reproducible results across software implementations, and it can lead to improved prior predictive assessments and other metrics. In the pages below, we show how opaque priors can arise from positive definite constraints, from sign indeterminacies, and from parameter order constraints. For each of these topics, we provide examples of the problem, discuss how the problem can lead to compounding problems in model assessment, and provide recommendations for avoiding the problem. Finally, we summarize and make general recommendations for practice.

\section{Positive definite constraints}
In Bayesian modeling, prior distributions for the covariance matrix often involve the inverse-Wishart (IW) distribution due to its conditional conjugacy. 
However, the IW distribution can be problematic because it assumes the same amount of prior information for the entire covariance matrix.  
To overcome this challenge, various strategies have been suggested for separately specifying priors on the variance and correlation parameters underlying the covariance matrix \citep{barmcc00}. These priors have been shown to perform better than the classical IW \citep{Alv14, Ari19, Hua13}. 
However, things become more complicated for matrix dimensions of three or more, because certain restrictions must be imposed to ensure positive definiteness of the matrix \citep{barmcc00, Dan99, Hua13, Rua15, Wei13}.

The situation becomes even more complicated when the covariance matrix has model-imposed constraints, which can arise in SEMs with correlated residuals or with across-group equality constraints. In these models, we cannot impose an IW (or other prior) on the full covariance matrix because those priors will not respect the constraints imposed by the model. An easily-implemented approach is to place priors on individual parameters within the covariance matrix (and we consider other approaches later). The problem is that, when we build a covariance matrix using these parameters, the resulting matrix will sometimes be non-positive definite. We elaborate below.

\subsection{Illustration}
\begin{figure}
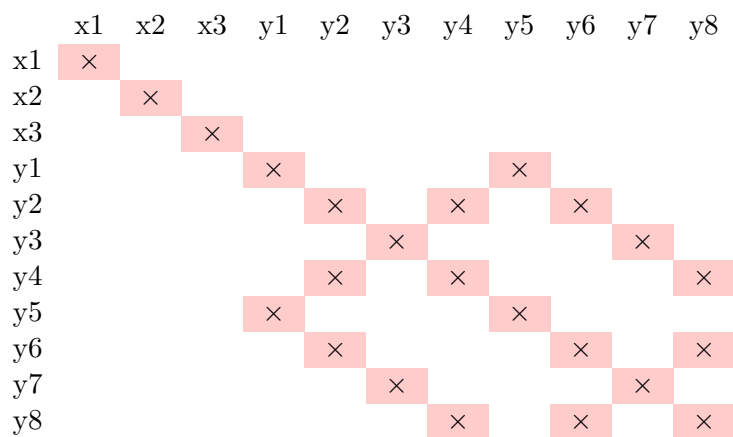

  \begin{center}
\begin{tabular}{lccccccccccc}
 & x1 & x2 & x3 & y1 & y2 & y3 & y4 & y5 & y6 & y7 & y8 \\
x1  & $\redc \times$ &  &  &  &  &  &  &  &  &  &  \\ 
x2  &  & $\redc \times$ &  &  &  &  &  &  &  &  &  \\ 
x3  &  &  & $\redc \times$ &  &  &  &  &  &  &  &  \\ 
y1  &   &  &  & $\redc \times$ &  &  &  & $\redc \times$ &  &  &  \\ 
y2  &   &  &  &  & $\redc \times$ &  & $\redc \times$ &  & $\redc \times$ &  &  \\ 
y3  &   &  &  &  &  & $\redc \times$ &  &  &  & $\redc \times$ &  \\ 
y4  &   &  &  &  & $\redc \times$ &  & $\redc \times$ &  &  &  & $\redc \times$ \\ 
y5  &   &  &  & $\redc \times$ &  &  &  & $\redc \times$ &  &  &  \\ 
y6  &   &  &  &  & $\redc \times$ &  &  &  & $\redc \times$ &  & $\redc \times$ \\ 
y7  &   &  &  &  &  & $\redc \times$ &  &  &  & $\redc \times$ &  \\ 
y8  &   &  &  &  &  &  & $\redc \times$ &  & $\redc \times$ &  & $\redc \times$ \\
\end{tabular}
  \end{center}
\caption{Political democracy model, structure of residual covariance matrix. Free parameters are marked by $\redc \times$.}
\label{fig:cstruc}
\end{figure}

An example comes from the popular ``Political Democracy'' model originally described by \cite{bol89}, shown below in \pkg{lavaan} syntax \citep{ros12}. Intended to describe countries' relationships between their levels of industrialization and democracy, the model contains four observed variables that are measured once during 1960 ($y1,...,y4$) and again during 1965 ($y5,...,y8$). The residuals of the 1960 variables are allowed to correlate with the residuals of the corresponding 1965 variables. Additionally, there exists a pair of similar variables collected during 1960 and again during 1965, leading to two more residual correlations. The full structure of the residual covariance matrix is shown in Figure~\ref{fig:cstruc}.

\newpage

\begin{knitrout}\footnotesize
\definecolor{shadecolor}{rgb}{0.969, 0.969, 0.969}\color{fgcolor}\begin{kframe}
\begin{alltt}
\hlstd{model} \hlkwb{<-} \hlstr{'
  # measurement model
    ind60 =~ x1 + x2 + x3
    dem60 =~ y1 + a*y2 + b*y3 + c*y4
    dem65 =~ y5 + a*y6 + b*y7 + c*y8
  # regressions
    dem60 ~ ind60
    dem65 ~ ind60 + dem60
  # residual correlations
    y1 ~~ y5
    y2 ~~ y4 + y6
    y3 ~~ y7
    y4 ~~ y8
    y6 ~~ y8
'}
\end{alltt}
\end{kframe}
\end{knitrout}

As described earlier, we could elect to put a univariate prior distribution on each variance (or standard deviation or precision) in this matrix, and also on the six residual correlations that are not fixed to zero. But this is problematic because the univariate priors on correlations can yield non-positive definite correlation matrices. The specific problem that occurs depends on the software package. For example, JAGS will stop as soon as it encounters a correlation matrix that is not positive definite. This means that univariate priors on correlations cannot often be used. On the other hand, Stan will report that a non-positive definite matrix was encountered, reject it, and continue sampling. We are left with univariate priors that are collectively constrained to be positive definite. If we consider only the space of positive definite correlation matrices under these priors, then the priors are usually more informative than we originally specified. That is, the prior distributions are opaque: the analyst specifies a set of priors that are different from the implied prior distributions, which must obey the constraint of positive definiteness.

\begin{figure}
\begin{knitrout}\footnotesize
\definecolor{shadecolor}{rgb}{0.969, 0.969, 0.969}\color{fgcolor}

{\centering \includegraphics[width=4in,height=3in]{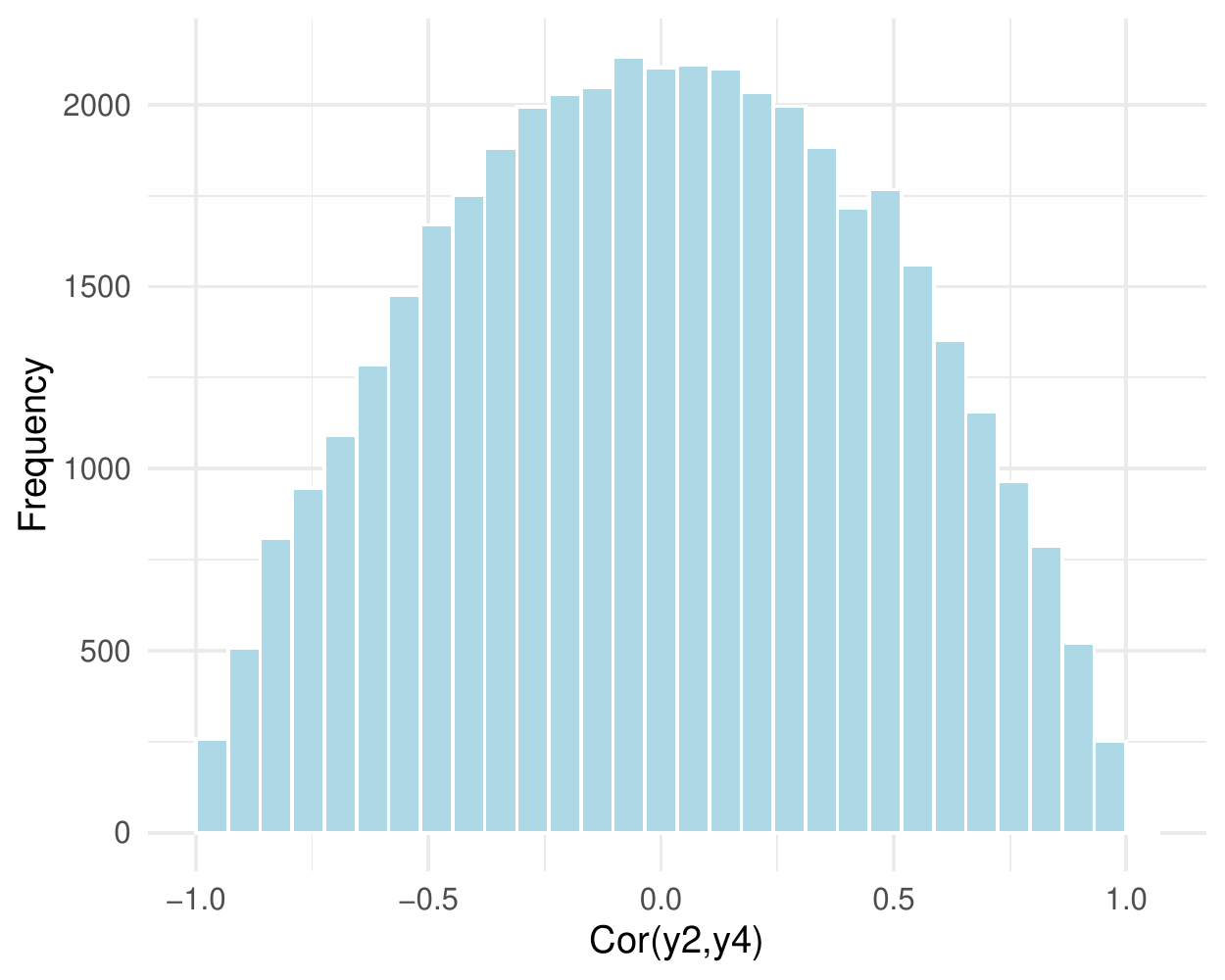} 

}

\end{knitrout}
\caption{Bollen model, implied prior distribution of a single correlation parameter after accounting for positive definite constraints.}
\label{fig:betapd}
\end{figure}

How can we characterize the implied prior distributions? We could simply generate thousands of correlation matrices from the prior, then discard matrices that are not positive definite, then visualize what is left. As an example of this, we specified a Uniform($-1,1$) prior for each free correlation in Figure~\ref{fig:cstruc}. We then generated 100,000 correlation matrices with the desired structure, discarding the 57,818 matrices that were not positive definite. Finally, we examined the distributions of the remaining matrices, with a histogram for a single correlation parameter appearing in Figure~\ref{fig:betapd}. We see that the resulting distribution is no longer uniform; it is approximately symmetric around 0, with more density near 0 than near $-1$ or $1$. This is because our uniform priors did not account for the fact that correlation matrices must be positive definite.

\begin{figure}
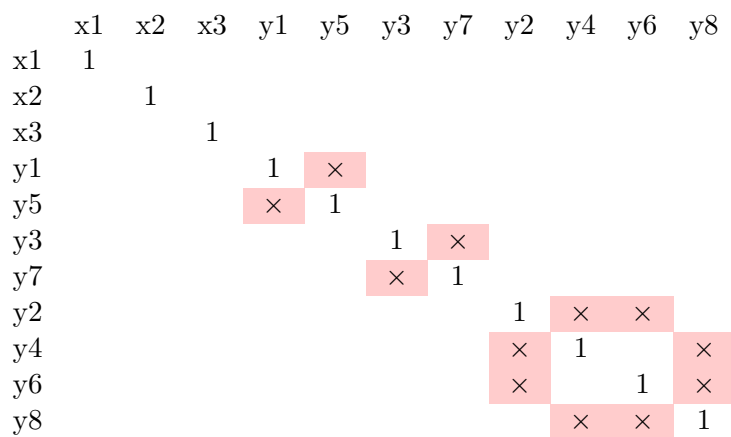

  \begin{center}
\begin{tabular}{lccccccccccc}
  & x1 & x2 & x3 & y1 & y5 & y3 & y7 & y2 & y4 & y6 & y8 \\
x1 & 1 &  &  &  &  &  &  &  &  &  &  \\
x2 &  & 1 &  &  &  &  &  &  &  &  &  \\
x3 &  &  & 1 &  &  &  &  &  &  &  &  \\
y1 &  &  &  & 1 & $\redc \times$ &  &  &  &  &  &  \\
y5 &  &  &  & $\redc \times$ & 1 &  &  &  &  &  &  \\
y3 &  &  &  &  &  & 1 & $\redc \times$ &  &  &  &  \\
y7 &  &  &  &  &  & $\redc \times$ & 1 &  &  &  &  \\
y2 &  &  &  &  &  &  &  & 1 & $\redc \times$ & $\redc \times$ &  \\
y4 &  &  &  &  &  &  &  & $\redc \times$ & 1 &  & $\redc \times$ \\
y6 &  &  &  &  &  &  &  & $\redc \times$ &  & 1 & $\redc \times$ \\
y8 &  &  &  &  &  &  &  &  & $\redc \times$ & $\redc \times$ & 1 \\
\end{tabular}
  \end{center}
  \caption{Political democracy model, structure of permuted correlation matrix. Free parameters are marked by $\redc \times$.}
\label{fig:pcstruc}
\end{figure}

\subsection{Positive Definiteness}
To technically describe the positive definite constraint here, we can simultaneously permute the rows and columns of the correlation matrix to obtain a block diagonal matrix. A block diagonal matrix is useful because the determinant of the full matrix is the product of determinants of each individual block within the matrix. This allows us to examine whether or not the full matrix is positive definite, by working with submatrices of smaller dimension.

The Cuthill-McKee algorithm \citep{cutmck69} allows us to automatically find an appropriate, block-diagonal permutation. After carrying out this algorithm \citep[via the \pkg{netprioR} package;][]{netprior} and permuting the rows and columns, we arrive at the matrix in Figure~\ref{fig:pcstruc}. This shows that, to keep the full matrix positive definite, we only need to worry about the $4\times4$ block in the lower right that involves $y2$, $y4$, $y6$, and $y8$. The priors on the $(y3,y7)$ and $(y1,y5)$ correlations have no influence on the positive definiteness of the full matrix (because a $2 \times 2$ matrix is positive definite for any correlation in $(-1,1)$), so that we can safely put a uniform prior on each of those correlations.

After applying traditional rules for computing matrix determinants, we can express the determinant of the $4\times4$ block of correlations as
\begin{equation}
  \label{eq:pdet}
  \det(\bm{R}_{(4\times4)}) = 1 + (r_1r_4 - r_2r_3)^2 - \sum_{i=1}^4 r_i^2.
\end{equation}
This shows analytically how the positive definite constraint of our correlation matrix influences our univariate priors. The univariate priors are collectively constrained by the requirement that Equation~\eqref{eq:pdet} be greater than 0. For MCMC algorithms that sample the correlations individually, each correlation is impacted by the current values of the other correlations. For example, if the current values of $r_1$, $r_2$, and $r_3$ are .9, .1, and .1, respectively, then $r_4$ must be less than .9 in order to maintain a positive definite matrix. But if the first three correlations assume different values, then it would be possible to observe values of $r_4$ above .9.

\subsection{Implications for Bayes factors}
To show how this issue becomes problematic for applied work, we consider the calculation of Bayes factors using the Bollen model. Imagine that we wish to know whether the two residual correlations involving $y2$ are necessary. To address this question, we could compute a Bayes factor comparing a model that includes those two residual correlations, to a model without those residual correlations. A computationally-cheap way to do this is the Savage-Dickey method \citep{diclie70,wag10}. This involves comparing the prior distributions of the two residual covariances to their posterior distributions, focusing on the point at which those covariances equal 0.

The Savage-Dickey method is slightly more complicated than usual because the two residual correlations in question influence the possible values that other residual correlations can take (due to positive definiteness). This changes the prior distributions on other residual correlations, as we move from a model with the two focal correlations freed to a model with the two focal correlations fixed. \cite{heck19} discusses that, in such a situation, we need to supply a correction term to the usual Savage-Dickey calculation that accounts for the change in prior distributions \citep[also see][]{verwas95}.

If we ignore (or do not realize) all of the above and set Uniform priors on each individual correlation, then the prior density of the two focal correlations at 0 equals $0.25$. This corresponds to a log-density of $-1.39$. In contrast, the true joint prior density on the two correlations of the full model (with the two focal correlations free, respecting the positive definite constraint) can be approximated using the positive definite correlation matrices that we randomly generated earlier. This approximations involves a density estimation method from the \pkg{ks} package in R \citep{kspack}. Our approximate joint density at 0 is now 0.46, corresponding to a log-density of $-0.78$. And we have an additional correction term to account for the fact that the priors for the $y4$--$y8$ and for the $y6$--$y8$ correlations are impacted by whether or not the $y2$ residual correlations are fixed to 0.

These evaluations lead to two separate Bayes factors for the model with correlations, relative to the model without correlations. Using our incorrect priors that do not account for positive definite constraints, we obtain a log-Bayes factor of 4.66 in favor of the model with correlations. Using our priors that do account for positive definite constraints, along with the correction from \cite{heck19}, we obtain a log-Bayes factor of 5.54 in favor of the model with correlations. Both of these Bayes factors provide support for the model with correlations, but possibly at different levels of evidence. For example, if we subscribe to the rules of thumb offered by \cite{kasraf95}, then these two Bayes factors lead us to conclude ``strong'' evidence using the incorrect prior calculation, and ``very strong'' evidence using the correct calculation.

We agree with you, the reader, that these cutoffs are arbitrary and that the Bayes factors do not differ by very much. But the point is that the Bayes factor systematically differ depending on how we compute prior densities. These differences will sometimes lead to different substantive conclusions in practice, with the easier computation (i.e., ignoring the positive definite constraint) being incorrect.


\subsection{Solutions}
While it was straightforward to visualize implied prior distributions in the Bollen model, the process becomes inefficient for correlation matrices whose dimension is larger than 3 or 4. For such correlation matrices, few of the randomly-generated matrices will be positive definite, and it will take a long time to obtain a sufficient number of positive definite matrices to describe the implied prior. Additionally, every unique structure of correlation matrix will have a unique positive definite constraint, similar to the one from Equation~\eqref{eq:pdet}. So we desire more general solutions that do not involve random generation of correlation matrices. We describe three solutions below that differ in complexity and in the extent to which they fully solve the problem.

\subsubsection{Informative priors}
The simplest, partial solution is to maintain the univariate priors on individual correlations, but make those priors informative around 0. For example, instead of placing Uniform priors on the correlations, we might use Beta distributed priors. The Beta distribution is typically defined for the interval $(0,1)$, but we can transform it so that the interval $(-1,1)$ is supported (and note that the Uniform is a special case of the Beta). To make this distribution more informative around 0, we can increase the value of both shape hyperparameters, for example Beta($5,5$). The implied prior distributions will then be closer to what the user specifies, because these informative priors will more often lead to positive definite correlation matrices. 
But this is not a full solution because, even with informative priors, we may still encounter non-positive definite correlation matrices. And it is not straightforward to predict when or how often this will happen. Additionally, depending on one's application, certain informative priors may be inappropriate.

\subsubsection{Priors on Cholesky decomposition} A more general solution to this problem comes from putting priors on the Cholesky decomposition of the correlation matrix, which is related to the \cite{ghomal21} approach for time series models. The Cholesky decomposition is advantageous because, to ensure that the correlation matrix stays positive definite, we only have to ensure that the diagonal elements of the Cholesky decomposition are positive.

For the Political Democracy model, the $4 \times 4$ covariance matrix from the bottom right corner of Figure~\ref{fig:pcstruc} has a Cholesky decomposition with structure
\begin{equation}
  \label{eq:pdchol}
  \begin{pmatrix}
    c_{11} & & & \\
    c_{21} & c_{22} & & \\
    c_{31} & -c_{21}c_{31}/c_{22} & c_{33} & \\
    0 & c_{42} & c_{43} & c_{44} \\
  \end{pmatrix},
\end{equation}
where the entries $\{c_{11}, c_{22}, c_{33}, c_{44}\}$ are constrained to be positive, while $\{c_{21}, c_{31}, c_{42}, c_{43}\}$ are unconstrained. Additionally, the entry in row 3, column 2 is fully determined by other entries. If we place gamma priors (say) on the diagonal entries and normal priors on the remaining $c$ variates, we can maintain the desired structure of the covariance matrix while also maintaining positive definiteness. 

A disadvantage of this approach is that the entries of the Cholesky decomposition do not necessarily have intuitive interpretations, so that it is difficult to set informative priors. Each diagonal entry is related to the portion of the corresponding variable's variance that cannot be accounted for by variables that occur further to the left of the matrix. Each off-diagonal entry is related to a partial correlation conditioned on variables further to the left of the matrix \citep[see][]{joe06,lkj09,poudan07}. This implies that the order of the variables matters. A further difficulty is that there does not appear to be an automatic way to obtain a Cholesky structure (like that of Equation~\eqref{eq:pdchol}) for arbitrarily-structured covariance matrices.

\subsubsection{Combining LKJ with Informative Priors} A final solution was recently described in a blog post by \cite{mar21} and can be implemented in Stan. This solution involves use of a prior distribution that is the product of (i) a Lewandowski-Kurowicka-Joe (LKJ) prior \citep{lkj09} on the full correlation matrix, and (ii) informative priors on individual entries of the correlation matrix. The resulting prior inherits the positive definiteness from (i) while also inheriting the informativeness from (ii). At the moment, it is not clear that this approach can be used to fix individual entries of the correlation matrix; instead, \cite{mar21} recommends highly-informative priors around the fixed value that is desired (similar to the notion of ``approximate zeros'' in a factor loading matrix). The highly-informative priors may be sufficient to replace some hard constraints, but they may also sometimes cause problems with convergence of the MCMC chains.

\section{Sign Indeterminacies}
We now turn to a problem that is more specific to SEM: sign indeterminacies of loading parameters. It is well known that, if we change the signs of all loadings, the SEM likelihood (usually) stays the same. To avoid this issue, SEM software typically ``prefers'' positive loadings through various aspects of implementation. First, for both Bayesian and frequentist models, the loadings' starting values are often set to positive numbers. Additionally, if a single loading is fixed for identification, it is almost always fixed to +1. This often leads other loadings towards positive values.

Especially when using software like JAGS or Stan, researchers commonly fix the latent variance to 1 and place truncated normal priors on the factor loadings, where the distributions are truncated from below at 0 \citep[e.g.,][]{cur10}. This forces all loadings to be positive and resolves sign indeterminacies in the model. But this solution is problematic because it does not allow for indicators with ``bad'' loadings (whose posterior distributions overlap with zero), and it does not allow for reversed indicators (whose valence is opposite that of other indicators). \cite{pee12} shows that, to achieve parameter identification of the likelihood, only one loading per latent variable must be sign restricted (with the latent variance being fixed to 1). Thus, fixing the signs of all loadings is overly restrictive from a parameter identification standpoint.

When a single loading per factor is fixed to 1 for identification, we should not need to fix the signs of any other loadings. If we instead fix the latent variance to 1 for identification, then an improved solution (over fixing all signs to positive) is to employ relabeling algorithms \citep[e.g.,][]{erocur17}. Under this approach, we allow factor loadings to flip between positive and negative values during MCMC estimation. Then, after model estimation, we change the signs of loadings depending upon the signs of some focal loading parameters (and, if the model includes factor correlations or factor regressions, we also may need to change those signs). This strategy leads to positive loading values, while allowing for the possibility that some loadings are negative.

The relabeling algorithms' preferences for positive loadings can conflict with reseachers' desires to use noninformative prior distributions for the loadings (say, Normal with a mean of 0 and a large variance). That is, the software's preference for positive loadings conflicts with the noninformative prior distributions, which state that both positive and negative loadings are equally likely. More generally, factor loadings are influenced by the model identification constraints \citep[e.g.,][]{bollil22}. To set meaningful prior distributions on factor loadings, researchers need to consider the specific identification constraints that will be used.

\subsection{Illustration}
To illustrate the interaction between sign indeterminacy and prior distributions, it is sufficient to consider the usual confirmatory factor model that is fit to the \cite{holswi39} data. We suspect that most people reading this far know the dataset, which contains scores on various tests of mental ability. We are focusing on the 3-factor model that is traditionally fit to the version of the data from \pkg{lavaan} \citep{ros12}, where each factor is associated with 3 observed variables.

The Holzinger-Swineford factor model has five types of model parameters: intercepts, loadings, factor standard deviations, factor correlations, and residual standard deviations. We assign true (``population'') values to all these parameters. Intercepts receive true values of 0, factor standard deviations receive true values of 1, factor correlations receive true values of 0, and residual standard deviations receive true values of 1. Finally and importantly, loadings receive true values of $-1$. 

Using these true values, we generated a dataset of 1,000 observations and re-fit the 3-factor model back to the data (where true values were treated as unknown). We used common, non-informative priors for the parameters of the estimated model, which are currently the defaults in \pkg{blavaan} \citep{merros18,merfit21}. 
\begin{align*}
  \text{Intercept} &\sim \text{N}(0, 1000) \\
  \text{Loading} &\sim \text{N}(0, 100) \\
  \text{Latent covariance matrix} &\sim \text{LKJ}(1) \\
  \text{Residual SD} &\sim \text{Gamma}(1,.5),
\end{align*}
where the Normal distributions are parameterized with variances, and where the LKJ prior is placed on the entire latent covariance matrix at once and respects the positive definite constraints described earlier. To identify the model, we fixed each latent variance to 1. In this case, \pkg{blavaan} uses a relabeling algorithm to handle sign indeterminacy. The estimation had three chains, 500 warmup samples per chain, and 1,000 posterior samples per chain.

\begin{figure}
\begin{knitrout}\footnotesize
\definecolor{shadecolor}{rgb}{0.969, 0.969, 0.969}\color{fgcolor}

{\centering \includegraphics[width=4.5in,height=4.5in]{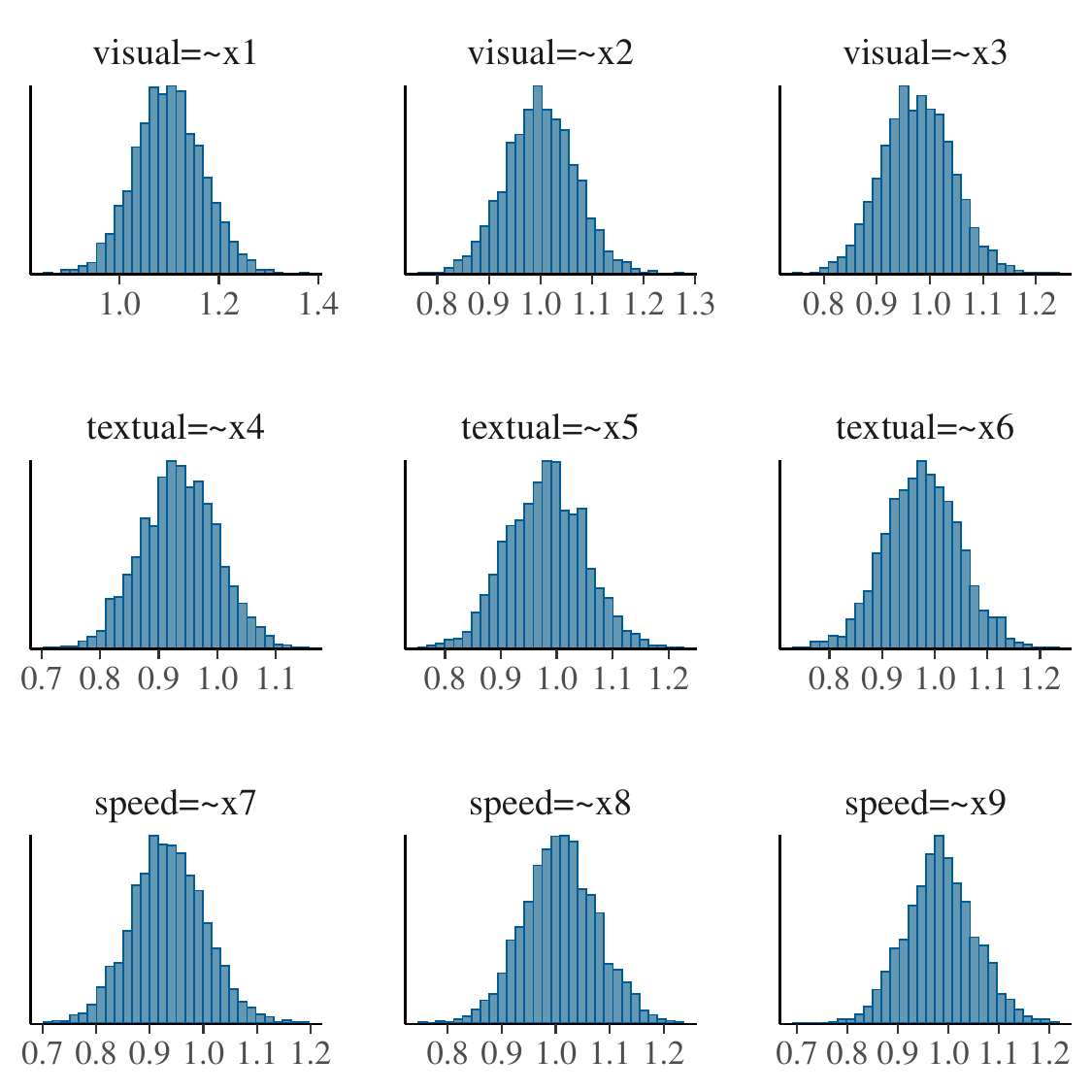} 

}

\end{knitrout}
\caption{Estimated posterior distributions of loadings, Holzinger-Swineford 3-factor model.}
\label{fig:signflip}
\end{figure}

The resulting posterior distributions of the loadings appear in Figure~\ref{fig:signflip}. These distributions are centered near $+1$, with the distributions being fully on the positive side of the space. So we have a situation where the true loading values were $-1$, we used noninformative priors that exhibit little influence on the posterior, and the posterior distributions of loading values are nowhere near the true values.

At this point, readers might object that this just illustrates sign indeterminacy. And posterior inferences about factor loadings are not generally impacted here. But the example highlights that, for loadings, a noninformative prior centered at 0 usually ignores the identification constraint that was chosen. That is, when employing noninformative priors, we are usually attempting to say that we have no idea about the loadings' values, or to avoid influencing the results of the model estimation. But the prior ignores the fact that (i) we typically fix a loading to be positive for identification, and (ii) observed variables are usually positively correlated, so that we can expect other loadings to be positive. So our original priors, which were intended to be \textit{noninformative}, may actually state that an implausible part of the parameter space is plausible \citep[also see][]{seaman_hidden_2012, gelman_prior_2017}. Further, as mentioned earlier, the prior distribution of relabelled factor loadings generally differs from the prior distribution of un-relabelled factor loadings, which can impact post-estimation computations that rely on evaluation of the prior.

\subsection{Implications for MCMC Validation}
Sign indeterminacy also complicates MCMC algorithm validation, which is used to ensure that MCMC samplers are working correctly. The MCMC validation process is difficult even without sign indeterminacy, because randomness is inherent in MCMC sampling. This means that we cannot simply examine whether the posterior means and standard deviations match other samplers to many decimal places. While it is possible to obtain analytic posterior distributions for certain models, analytic results are the exception instead of the rule for models estimated via MCMC.

{\em Simulation-based calibration} \citep[SBC;][]{modmoo22,talts2018} is an MCMC validation method that has received recent attention and implementation. Given a model of interest (including likelihood and priors), simulation-based calibration can be described in four steps.
\begin{enumerate}
\item Generate many sets of parameter values from the prior distribution.
\item For each set of parameters from Step 1, generate an artificial dataset.
\item Fit the model of interest to each artificial dataset from Step 2.
\item Examine whether the resulting posterior distributions look like the prior distribution.
\end{enumerate}
If the MCMC sampler is working correctly, then the posteriors from Step 3 should look like the priors from which we started. We can graphically examine this idea by comparing the posterior means from Step 3 to the parameter values from Step 1; we should see an identity line when plotting the parameter values against the posterior means.

Using the same model from the previous section, we used the \pkg{SBC} package \citep{sbc} to conduct simulation-based calibration under two sets of priors, with 1000 simulated data sets each. Set 1 was exactly the same as the noninformative priors from the previous section. Set 2 was also similar to the previous section, differing only in the priors for the loadings: instead of Normal(0, 100), the priors for loadings were Normal(1, 1/16) (where the second number is a variance).
This is an informative prior reflecting the belief that all loadings should be similar to one another (recall that a single loading is being fixed to 1 for identification).

\begin{figure}
\begin{knitrout}\footnotesize
\definecolor{shadecolor}{rgb}{0.969, 0.969, 0.969}\color{fgcolor}

{\centering \includegraphics[width=6in,height=6in]{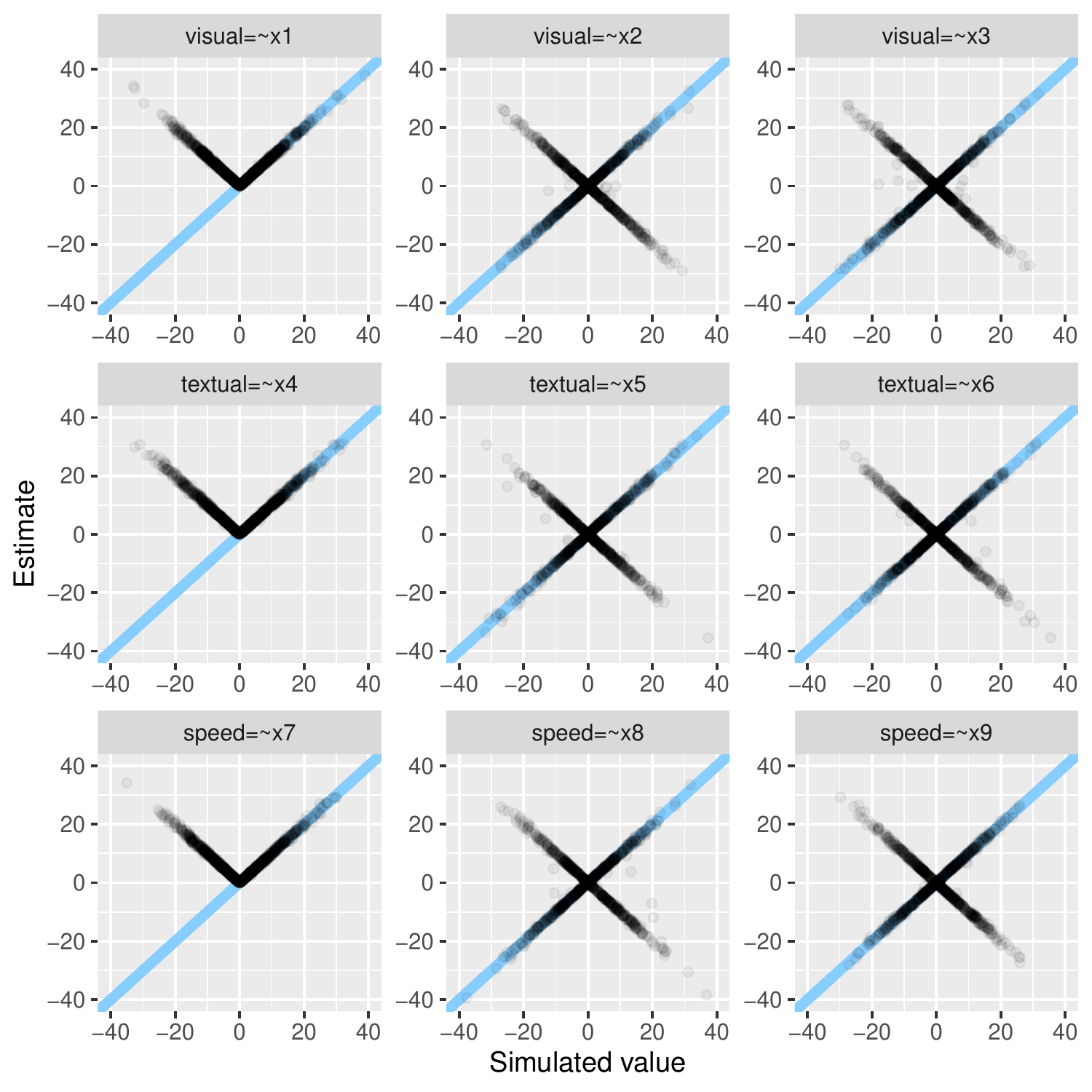} 

}

\end{knitrout}
\caption{Prior parameter values versus posterior means for N(0,100) priors on loadings.}
\label{fig:sbcni}
\end{figure}

\begin{figure}
\begin{knitrout}\footnotesize
\definecolor{shadecolor}{rgb}{0.969, 0.969, 0.969}\color{fgcolor}

{\centering \includegraphics[width=6in,height=6in]{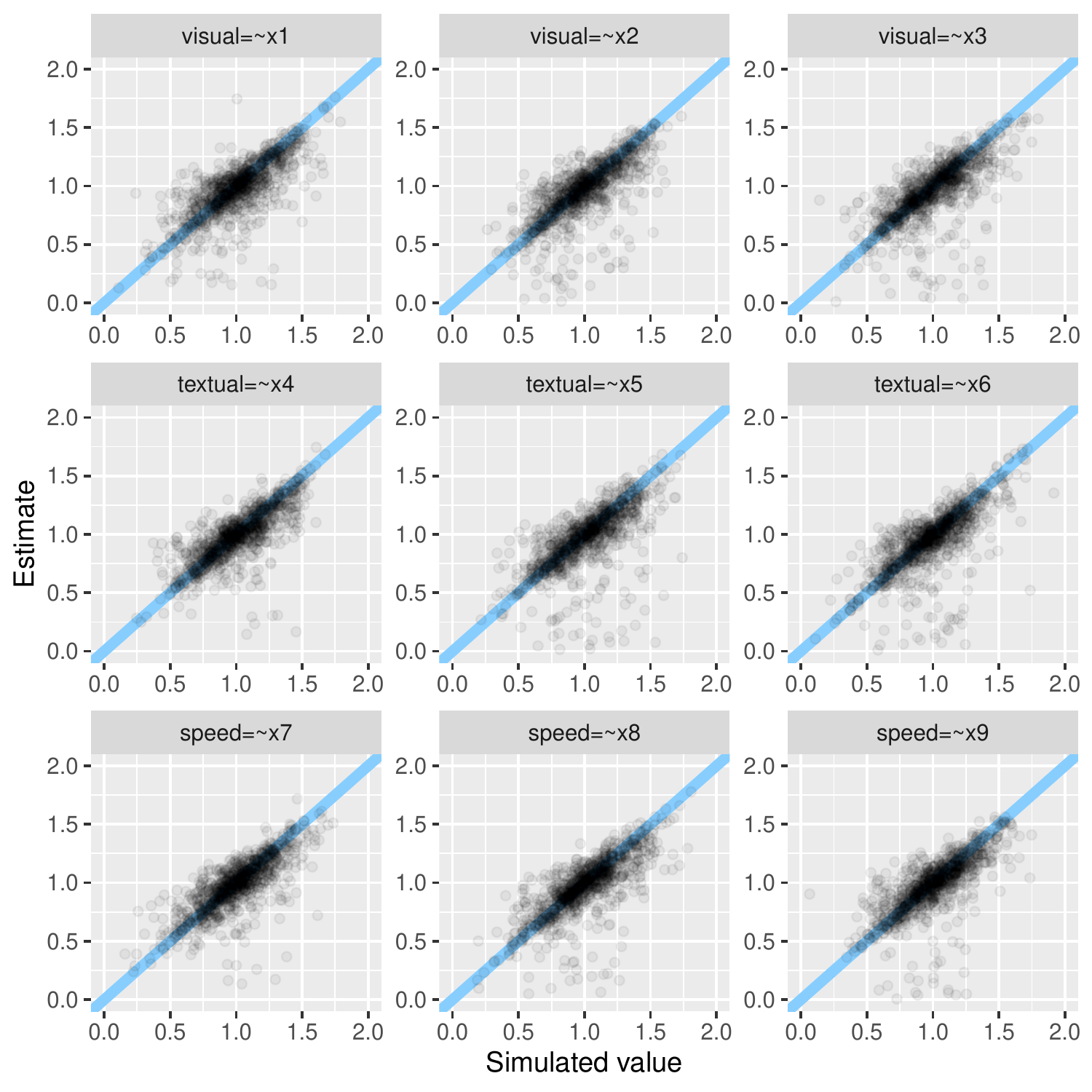} 

}

\end{knitrout}
\caption{Prior parameter values versus posterior means for N(1,$\frac{1}{16}$) priors on loadings.}
\label{fig:sbcinf}
\end{figure}

Results for the Set 1 and Set 2 priors are shown in Figures~\ref{fig:sbcni} and~\ref{fig:sbcinf}, respectively. In both figures, the parameter values simulated from the prior are on the x-axis, and the posterior means estimated from the artificial data are on the y-axis. Each panel represents a factor loading (there are 6 free loadings in the model). Each point represents a replication, and ``correct'' MCMC algorithms should lead to points that are crowded around the blue diagonal line.

Figure~\ref{fig:sbcni} shows that, for noninformative priors on factor loadings, the points follow a V or an X pattern. The V pattern occurs for loadings that are constrained to be positive during estimation, while the X pattern occurs for the remaining loadings. This means that the estimation algorithm often recovers the true parameter value, but it also often flips the sign of the parameter value. While this is not a problem if we realize the sign indeterminacy issue, it is easy to overlook in the context of simulation-based calibration. For example, typical SBC summaries involve cumulative distributions, which would lead us to conclude that there are problems with the MCMC algorithm, and which would not provide clear clues about sign indeterminacy. See \cite{merfit21} for some discussion of similar analyses.

Figure~\ref{fig:sbcinf}, on the other hand, is closer to what we would hope to see from a correct MCMC algorithm. In this figure, the points generally form a cloud around the blue diagonal line, implying that the posterior means match the parameter values that generated the data (note that the axis limits have changed, as compared to the previous figure). The informative priors on factor loadings work similarly to use of truncated normal priors, where the signs of the loadings were generally restricted to be positive. But informative priors differ from the truncated normals in that they allow for the possibility of negative loadings. 

\subsection{Solutions}
If researchers are aware of how their software handles sign indeterminacy, then they can potentially avoid the issues described here and successfully employ noninformative prior distributions. Barring that, we advise researchers to explicitly consider the loadings' expected signs when setting prior distributions. In many models, we expect that all the observed variables corresponding to a factor will have the same direction of relationship with that factor. Additionally, a single loading is often set to 1 for identification. In a situation like this, it is often reasonable to place priors on the free loadings that have a mean of 1 and a standard deviation of, say, .5. These priors look very informative at first glance, as compared to, say, a Normal prior with a mean of 0 and a variance of 10,000. But the suggested priors better represent what the researcher knows about the signs of the loadings, combined with the fact that some loadings are being fixed to 1.

Instead of fixing a single loading to 1 for identification, researchers may fix the latent variance to 1. In this case, as described by \cite{pee12}, one loading per latent variable must be constrained to be positive (or negative) in order to achieve identification. If that loading has a Normal prior whose support includes the negative (positive) reals, then we again face a conflict where our stated prior is not the implied prior. The implied prior for the sign-constrained loading may become truncated normal or truncated $t$, depending on other details of the MCMC implementation \citep[see][]{ghodun09}, while the priors on remaining loadings are as stated. While a truncated distribution arises from the identification constraint here, we think that truncated priors on all loadings should be avoided, because they prevent the loadings' posterior distributions from overlapping with zero. The relabeling strategies described by \cite{erocur17} should be preferred.

There are alternative prior distributions that avoid the issue and/or that make it easier to specify informative prior distributions, though they are not readily available in popular software. \cite{gramer22} studied prior distributions on ratios of factor loadings, as well as prior specification under effect coding \citep{litsle06}. Ratios of loadings avoid the sign-switching issue (e.g., the ratio of two negative loadings remains positive), and effect coding can make it easier to specify informative prior distributions.
Additionally, in the case of exploratory factor analysis, \cite{heaps22} recently proposed a prior distribution for the common factor covariance matrix. That is, factor analysis models typically imply a covariance matrix of the form $\bm{\Lambda} \bm{\Lambda}^\prime + \bm{\Psi}$, where $\bm{\Lambda}$ is the factor loading matrix with many fewer columns than rows. \cite{heaps22} proposes to place a matrix normal prior on $\bm{\Lambda} \bm{\Lambda}^\prime$, which encodes researcher knowledge about shared variation in the observed variables. The term $\bm{\Lambda} \bm{\Lambda}^\prime$ remains invariant across factor loadings' signs and rotations, so it avoids the issues described here. The proposed priors can also shrink sets of loadings towards zero, which is similar to confirmatory factor analysis. It remains to be seen how this prior could be translated to more general SEMs, where the model-implied covariance matrix becomes more complex.

\section{Order Constraints}
Finally, we describe prior distributions for order-constrained parameters, which are commonly seen in SEMs for ordinal variables with more than two categories. For these models, there exist threshold parameters that chop each underlying continuous variable into observed, ordered categories. The threshold parameters must be ordered so that they correspond to the ordering of the observed variables. For example, the lowest threshold chops off the lowest category, the second threshold chops off the bottom two categories, and so on.

The prior distributions for threshold parameters are often opaque, because the priors that researchers specify often have no order constraints. This is commonly done to improve the software's ease of use: researchers are accustomed to setting univariate Normal priors on individual parameters, and the priors with order constraints typically do not have simple forms. But the software always imposes order constraints here, which changes the prior distribution in various manners. Researchers often do not realize that anything happened, which may be especially problematic when setting informative priors.

\subsection{Example}
Say that a researcher fits a factor analysis model to a set of 4-category ordinal variables, and that she specifies a Normal(0,5) prior on all threshold parameters in the model. Because there are four categories per variable, we require three order-constrained thresholds per variable. We wish to know what these priors look like, after accounting for the order constraints.

We consider two ways that we could translate Normal(0,5) priors to three ordered parameters \citep[also see][]{padmor23}. First, we could imagine drawing three separate variates from the normal distribution, then ordering the variates to obtain ordered thresholds. Second, we could imagine drawing the first (lowest) threshold from a Normal(0,5), then adding a Lognormal(0,5) variate to that threshold in order to obtain the second threshold. Once we have obtained the second threshold, we could add another Lognormal variate to obtain the third threshold. Lognormal variables can only take positive values, so we are guaranteed to have an ordered set of thresholds under this approach.

For both of these translations, the threshold parameters' prior distributions differ from the Normal(0,5) distribution that the researcher originally declared. We expand on this point below, separately for the two methods.

\subsubsection{Reordering}
\begin{figure}
\begin{knitrout}\footnotesize
\definecolor{shadecolor}{rgb}{0.969, 0.969, 0.969}\color{fgcolor}

{\centering \includegraphics[width=5in,height=3in]{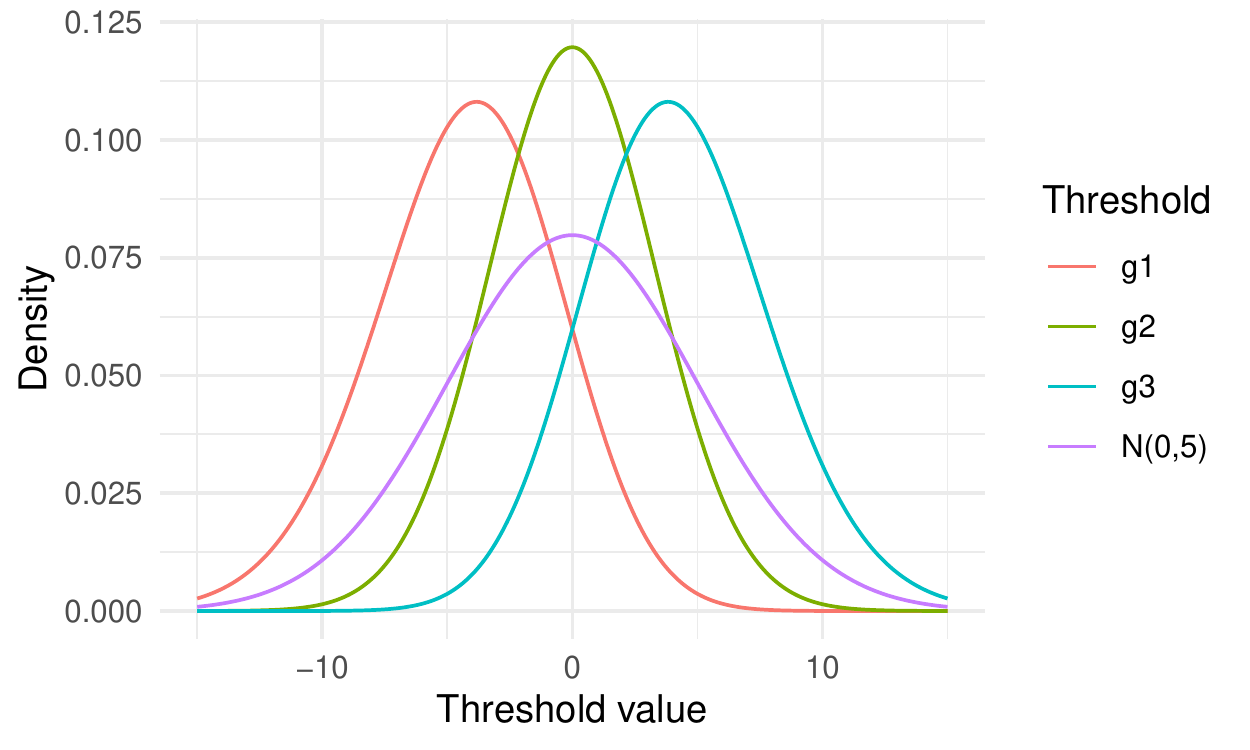} 

}

\end{knitrout}
\caption{Priors of threshold parameters whose stated priors are Normal(0,5), as implied by order constraints.}
\label{fig:impthres}
\end{figure}

When we draw three values from the Normal distribution and then order them, the act of ordering influences the resulting prior distributions. The specific distributions can be described via statistical theory on order statistics. For our example, the Normal(0,5) priors translate into the following probability density functions (pdfs) for individual thresholds:
\begin{align*}
  p(g_1) &= 3 \times \phi(g_1/5) \times [1 - \Phi(g_1/5)]^2 \\
  p(g_2) &= 6 \times \phi(g_2/5)  \times \Phi(g_2/5) \times [1 - \Phi(g_1/5)] \\
  p(g_3) &= 3 \times \phi(g_2/5)  \times \Phi(g_2/5)^2,
\end{align*}
where $\phi()$ is the standard normal pdf and $\Phi()$ is the standard normal cdf. These distributions are visualized in Figure~\ref{fig:impthres}, with the Normal(0,5) distribution overlayed for comparison. The figure shows that the prior for the lower threshold ($g_1$) is centered below 0, while the prior for the upper threshold ($g_3$) is centered above 0. None of the three distributions matches the Normal(0,5), despite the fact that the researcher declared a Normal(0,5) prior for all three parameters.

\subsubsection{Lognormals}
The \pkg{blavaan} package uses the lognormal approach mentioned previously. The Normal(0,5) prior goes on the lowest threshold parameter per item. Then, Normal(0,5) priors are placed on log-differences between subsequent parameters. For example, considering the three thresholds from the example, we would have
\begin{align*}
  g_1 &\sim \text{Normal}(0,5) \\
  \log(g_2 - g_1) &\sim \text{Normal}(0,5) \\
  \log(g_3 - g_2) &\sim \text{Normal}(0,5),
\end{align*}
where we could alternatively say that the differences between thresholds follow Lognormal priors.

Just like the previous section, the above priors can be translated to priors on individual thresholds. It is obvious that the prior for $g_1$ continues to be the stated prior, which is Normal(0,5). But the priors for $g_2$ and $g_3$ involve the sum of a normal distribution and Lognormal distribution(s). The resulting distributions can be written as
\begin{align*}
  p(g_2) &= \displaystyle \int_{-\infty}^{g_2} \phi((g_1 - \mu)/\sigma) \times \text{logN}(g_2 - g_1, \mu, \sigma) \partial g_1 \\
  p(g_3) &= \displaystyle \int_{-\infty}^{g_3} \int_{-\infty}^{g_2} \phi((g_1 - \mu)/\sigma) \times \text{logN}(g_2 - g_1, \mu, \sigma) \times \text{logN}(g_3 - g_2, \mu, \sigma) \partial g_1 \partial g_2 , \\
\end{align*}
where $\text{logN}(x, \mu, \sigma)$ is the density function of the Lognormal distribution with mean $\mu$ and standard deviation $\sigma$ (both on the log scale), evaluated at $x$.

\begin{figure}
\begin{knitrout}\footnotesize
\definecolor{shadecolor}{rgb}{0.969, 0.969, 0.969}\color{fgcolor}

{\centering \includegraphics[width=5in,height=3in]{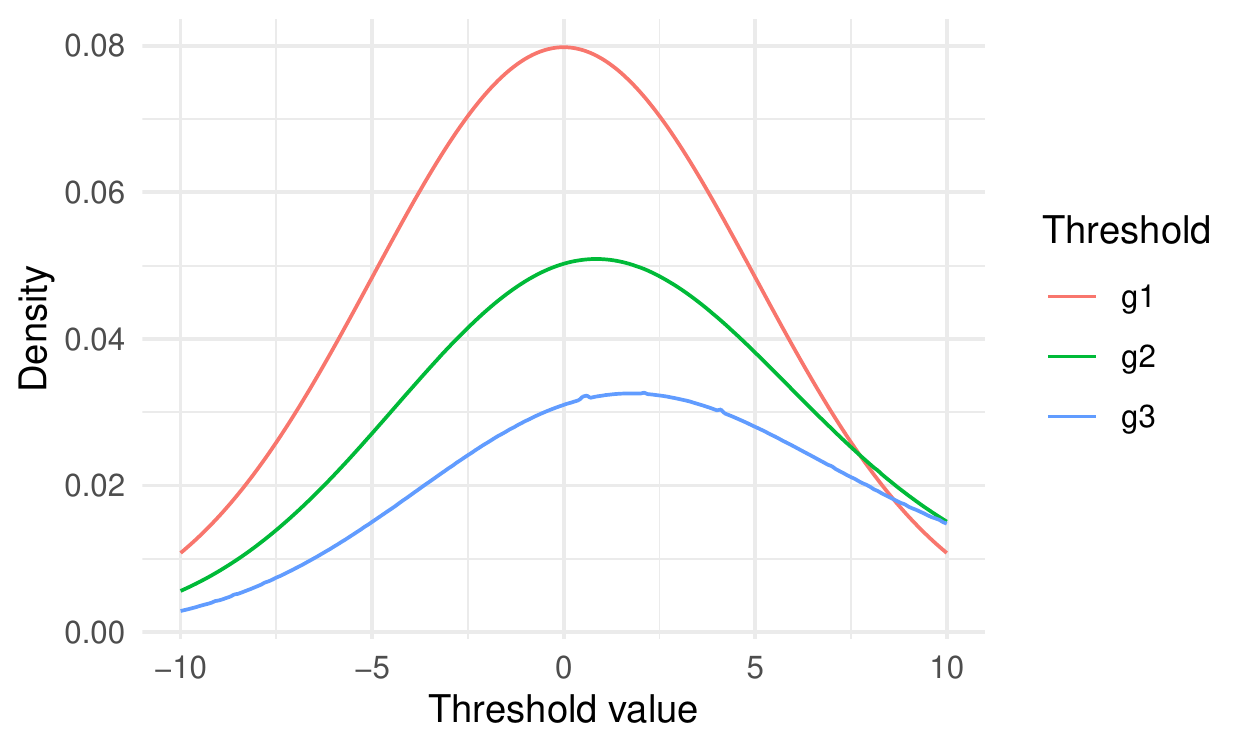} 

}

\end{knitrout}
\caption{Priors of threshold parameters whose stated priors are Normal(0,5), as implied by placing priors on log-differences.}
\label{fig:blathres}
\end{figure}

While the priors for $g_2$ and $g_3$ do not have nice forms, we can numerically approximate the integrals to visualize them. These distributions are shown in Figure~\ref{fig:blathres}, which is arranged similarly to Figure~\ref{fig:impthres}. As stated before, the lowest threshold, $g_1$, has the stated Normal(0,5) distribution. The peaks of the distributions of the remaining thresholds are closer to that of $g_1$, as compared to the distributions in Figure~\ref{fig:impthres}. The distributions of $g_2$ and $g_3$ are also skewed to the right, reflecting the fact that we are adding a positive variate to the distribution of $g_1$.

\subsection{Solution}
Unlike the previous issues with positive definite constraints and sign constraints, researchers do not have to consider changing their priors in order to address order constraints. The main solution is to be aware of the fact that, if one places univariate prior distributions on a set of order-constrained parameters, then some translation will take place to ensure that the parameters are ordered correctly. And this translation will influence the implied prior distribution of each parameter. It is worthwhile to understand how this is handled by one's software, especially for prior predictive assessments, Bayes factor calculation, and simulation-based calibration methods.

\section{General Discussion}
In this paper, we considered the idea of {\em opaque} prior distributions in Bayesian models with latent variables, where model estimation and prior specification interact to potentially yield unexpected results. The problem arises from the fact that software implementations must respect various model constraints, while researchers' prior specifications do not always respect the same constraints. This leads researchers to declare one set of prior distributions, which can imply a different set of prior distributions depending on the software implementation. The issue is particularly problematic for software verification, for comparing model results across different pieces of software, and for computing metrics that explictly involve evaluation of the prior distribution.

The three issues that we considered were (i) positive definite constraints on model covariance matrices; (ii) sign indeterminacy and constraints used to identify model parameters (typically factor loadings); and (iii) order constraints on subsets of model parameters (typically thresholds/intercepts). These issues occur with different frequencies, with (ii) and (iii) occurring more often than (i). To expand on this, issue (iii) occurs for most models that have ordinal variables with more than two categories, issue (ii) occurs for most measurement models (with free loadings), and issue (i) occurs for models with residual covariances, or other combinations of fixed and free covariances. We could have a worst-case scenario, such as a multiple group model with ordinal variables and across-group parameter constraints, where all three issues occur at once.

To avoid problems associated with opaque priors, we offer the following recommendations for practice:
\begin{enumerate}
\item If one's model involves covariance matrices without parameter constraints, use a single prior for the full covariance (or correlation) matrix (LKJ, inverse Wishart, etc).
\item If one's model involves a covariance matrix with parameter constraints, consider putting a prior on the Cholesky decomposition, or use matrix identities to see whether the full matrix can be broken into blocks that are easier to handle. If these are unavailable, use informative priors on the correlations that place more density close to 0.
\item For factor loadings, consider the expected direction of the relationship between each observed variable and the corresponding latent variable(s), along with the loading identification constraints. Use priors that place most density in this expected direction.
\item Be aware of how order constraints influence priors for thresholds, especially if one is doing model assessments that directly involve prior evaluation.
\end{enumerate}
Out of these recommendations, the priors on constrained covariance matrices are most difficult to handle. Future work could make it easier for researchers to place reasonable priors on constrained covariance matrices.

Importantly, the issue of opaque priors does {\em not} mean that all results associated with affected models are wrong. In the presence of opaque priors, we can still obtain accurate posterior summaries, including model information criteria (like WAIC and LOO) and some fit indices \citep[e.g.,]{garnier_adapting_2020}. On the other hand, the prior distributions that researchers describe in their papers may be incorrect, as will prior predictive checks and other model summaries that directly rely on prior distribution evaluation, such as Bayes factors. Researchers should especially be careful about applying Bayes factor computation strategies \citep[e.g.,][]{grosin20} to latent variable models, to ensure that they are evaluating the correct priors.

Opaque priors are vaguely similar to applied modeling of ordinal variables \citep[e.g.,][]{burvuo19,lidkru18}, where researchers ignore the fact that they have ordinal variables, treat them as continuous, and sometimes obtain reasonable results. Similarly, researchers can ignore the fact that they have opaque priors, estimate their model, and sometimes obtain reasonable results. In both cases, it is difficult to predict exactly when the results will be reasonable and when they will not. And ignoring the issues do not make them disappear.

We conclude by considering that some non-Bayesian researchers may find this paper appealing, because they can use it to justify phrases like ``Bayesian methods are difficult to use.'' We agree that priors present extra complications that do not exist for other methods, but we find the extra complications to be worthwhile. In our experience, wrestling with prior distributions can lead to a deeper, more sober understanding of one's model and how it interacts with data. This understanding might be achieved via other, non-Bayesian routes, but it will require the time and effort that Bayesians devote to prior distributions.

\section*{Computational Details}

All results were obtained using the \proglang{R}~system for statistical computing \citep{rprog},
version~4.3.1, with major reliance on packages \pkg{blavaan} \citep{merfit21}, \pkg{rstan} \citep{rstan}, and \pkg{SBC} \citep{sbc}. 


\bibliography{refs}

\end{document}